\documentclass[3p,twocolumn]{elsarticle}

\usepackage[first,bottomafter,timestamp]{}
\usepackage{mathrsfs}
\usepackage{hyperref,graphicx,amsfonts,amssymb,amsthm,amsmath,psfrag}
\newcommand{\be}[1]{\begin{equation} \centering \label{#1}}
\newcommand{\ee}{\end{equation}}
\newcommand{\ba}[1]{\begin{eqnarray} \centering \label{#1}}
\newcommand{\ea}{\end{eqnarray}}

\begin{document}

\begin{frontmatter}
\title{Higgs-Dilaton cosmology: Are there extra relativistic species?}

\author[mad]{Juan Garc\'ia-Bellido}
\ead{juan.garciabellido@uam.es}

\author[ch]{Javier Rubio}
\ead{javier.rubio@epfl.ch}  

\author[ch]{Mikhail Shaposhnikov} 
\ead{mikhail.shaposhnikov@epfl.ch}

\address[mad]{ Instituto de F\'{\i}sica Te\'orica CSIC-UAM, 
Universidad Aut\'onoma de Madrid, Cantoblanco, 28049 Madrid, Spain }

\address[ch]{
  Institut de Th\'eorie des Ph\'enom\`enes Physiques,
 \'Ecole Polytechnique F\'ed\'erale de Lausanne,
  CH-1015 Lausanne, Switzerland}

\begin{abstract}
Recent analyses of cosmological data suggest the presence of an extra
relativistic component beyond the Standard Model content. The
Higgs-Dilaton cosmological model predicts the existence of a massless
particle - the dilaton - associated with the spontaneous symmetry breaking
of scale invariance and undetectable by any accelerator experiment. 
Its ultrarelativistic character makes it a suitable candidate for
contributing to the effective number of light degrees of freedom in
the Universe. In this Letter we analyze the dilaton production at the
(p)reheating stage right after inflation and conclude that no extra
relativistic degrees of freedom beyond those already present in the
Standard Model are expected within the simplest Higgs-Dilaton
scenario.  The elusive dilaton remains thus essentially undetectable
by any particle physics experiment or cosmological observation.
\end{abstract}


\end{frontmatter}
%\maketitle
 %%%%%%%%%%%%%%%%%%%%%%%%%%%%%
\section{Introduction}
%%%%%%%%%%%%%%%%%%%%%%%%%%%%%
Cosmology is entering in a precision era where the interplay with
particle physics is becoming more and more important.  A noteworthy
example is the effective number of light degrees of freedom
appea\-ring in the different extensions of the Standard Model (SM). 
Any extra radiation component in the Universe is usually parametrized,
independently of its statistics, in terms of an effective number of
neutrino species, $N_\textrm{eff}=N_\textrm{eff}^\textrm{SM}+\Delta
N_{\textrm{eff}}$ \cite{Steigman:1977kc}, where
$N_\textrm{eff}^\textrm{SM}$ stands for the number of active neutrinos
in the SM\footnote{In the standard cosmological model with three
neutrino flavors and zero chemical potential we have
$N_\textrm{eff}^\textrm{SM}=3$ at BBN, and
$N_\textrm{eff}^\textrm{SM}=3.046$ at CMB. The small excess in the
last case with respect to the LEP result  \cite{Beringer:1900zz} is
due to the entropy transfer between neutrino species and the thermal
bath during  electron-positron annihilation
\cite{Lesgourgues:2006nd}.}. 

The strongest constraints on the effective number of neutrino species
come from Big Bang Nucleosynthesis (BBN). A non-standard value of
$N_\textrm{eff}$ increases the expansion rate, which  results on an
enhancement of the primordial helium abundance. Assuming zero lepton
asymmetry, the number of effective degrees of freedom at BBN turns out
to be $N_\textrm{eff}=3.71^{+0.47}_{-0.45}$ ($68 \%$ C.L.)
\cite{Steigman:2012ve}.  Note that, although the existence of extra
species is somehow favored, the obtained value is still compatible
with the SM prediction within the $95\%$ C.L. 

Some constraints on $N_\textrm{eff}$ can be also obtained from the
analysis of the Cosmic Microwave Background (CMB), although the
current limits are significantly weaker than those of BBN. The
combined analysis of WMAP7 results,  Hubble constant measurements and
baryon acoustic oscillations \cite{Komatsu:2010fb} provides a value
$N_\textrm{eff}=4.34^{+0.86}_{-0.88}$ ($68 \%$ C.L.). Similar and
complementary results for smaller CMB scales have been also reported
by the Atacama Cosmo\-logy Telescope \cite{Dunkley:2010ge} and the
South Pole Telescope \cite{Keisler:2011aw}.  It is interesting to
notice the dependence of the e\-ffective number of neutrino species on
the priors considered in the different Bayesian analysis existing in
the literature. While in some references the SM value, $N^{\rm
SM}_{\textrm{eff}}=3$, is ruled out at 95\% C.L.
\cite{GonzalezGarcia:2010un,Hou:2011ec},  in others, such as
\cite{Reid:2009nq}, it is not. Besides, if the helium abundance
obtained from CMB measures is taken into account, together with the
most precise primordial deuterium abundance \cite{Pettini:2012ph}, the
BBN result becomes perfectly consistent with the SM one at the
$2\sigma$ level, $N_\textrm{eff}=3.22\pm 0.55$ \cite{Steigman:2012ve}.
The number of extra degrees of freedom is therefore an open question
to be solved by the Planck satellite, which is expected to determine $
N_\textrm{eff}$ with an accuracy of $\sim 0.3$ at $2\sigma$
\cite{Ichikawa:2008pz}, breaking thereby the degeneracies with nonzero
neutrino masses and dynamical dark energy \cite{Hamann:2010pw}.

In order to account for the apparent radiation excess one can consider
several possibilities. It could be, for instance, the indication of an
extra sterile neutrino \cite{Hamann:2010bk,Hannestad:2012ky}, of relic
gravitational waves \cite{Giovannini:2002qw}, or  arise from other
exotic possibilities such as a decaying particle
\cite{Ichikawa:2007jv,Zhang:2007zzh,Fischler:2010xz}, the
interaction between dark energy and dark matter \cite{Bjaelde:2012wi} or the reheating of the
 neutrino thermal bath \cite{Boehm:2012gr}.
In this Le\-tter we will consider a different possibility within the
minimalistic framework of Higgs-Dilaton cosmology 
\cite{Shaposhnikov:2008xb,GarciaBellido:2011de,Shaposhnikov:2008xi}. 
This  constitutes an extension of the Higgs inflation idea
\cite{Bezrukov:2007ep}, where the Standard Model Higgs doublet $H$ is
non-minimally coupled to gravity.  The novel ingredient  of
Higgs-Dilaton cosmology is the invariance of the action under scale
transformations. This extra symmetry leads to the absence of any
dimensional parameters or scales\footnote{In particular it forbids the
appearance of a cosmological constant term in the action. In
Higgs-Dilaton cosmology,  the late dark energy dominated period of the
Universe is reco\-vered, at the level of the equations of motion, by
repla\-cing General Relativity with Unimodular Gravity. However, both
the inflationary and prehea\-ting stages considered in this Letter take
place in field space regions where the dark energy contribution is
completely ne\-gligible. We will thus omit  this point here. The
reader is referred to Ref. \cite{GarciaBellido:2011de} for details
about the phenomenological consequences of  Unimodular Gravity in the
Higgs-Dilaton scenario.}. The simplest phenomenologically viable
theory of this kind requires the existence of a new scalar singlet
under the SM gauge group  \cite{Shaposhnikov:2008xb}, the dilaton
$\chi$, non-minimally coupled to gravity. It corresponds to the
Goldstone boson associated with the spontaneous symmetry breaking of
scale invariance and it is therefore massless. This property makes it
a potential candidate for contributing to the effective number of
relativistic degrees of freedom at BBN and recombination. Indeed, this
cosmological test seems to be the only available probe for determining
the existence of the dilaton particle. The coupling between the dilaton
and all the SM fields (apart from the Higgs) is forbidden by quantum
numbers, which, together with the Goldstone boson nature of this
particle, excludes the possibility of a direct detection in an
accelerator experiment \cite{Shaposhnikov:2008xb}. 

In this Letter we study the (p)reheating stage in Higgs-Dilaton
cosmology, paying special attention to the dilaton production. This
Letter is organized as follows. In Sections \ref{sec:model} and
\ref{sec:gaugesector} we review the Higgs-Dilaton inflationary model
and show that, given the hierarchical structure of the non-minimal
couplings to gravity, the production of SM particles takes place, up
to some small corrections, as in the simplest Higgs inflationary
scenario \cite{Bezrukov:2008ut,GarciaBellido:2008ab,Bezrukov:2011sz}.
The difference between the two models is described in Section
\ref{sec:dilatonproc}, where we compute the dilaton production,
compare it with the total energy density of SM particles at the end of
the preheating stage and determine its contribution to the effective
number of relativistic degrees of freedom. The  conclusions are
presented in Section \ref{sec:conclusions}.

%%%%%%%%%%%%%%%%%%%%%%%%%%%%%
\section{Higgs-Dilaton inflation}\label{sec:model}
%%%%%%%%%%%%%%%%%%%%%%%%%%%%%

We start by reviewing the Higgs-Dilaton model
\cite{Shaposhnikov:2008xb,GarciaBellido:2011de}. In the unitary gauge 
$H^T=(0,h/\sqrt{2})$, it is described by the following Lagrangian
density
\be{general-theory}
\frac{\mathscr L}{\sqrt{-g}}=
\frac{1}{2}(\xi_\chi \chi^2 +\xi_h h^2)R-
\frac{1}{2}\left(\partial \chi\right)^2-U(\chi,h) \ ,
\ee
where we have omitted the part of the SM Lagrangian not involving the
Higgs potential, $\mathscr L_{\text{SM}[\lambda\rightarrow0]}$. The
values of the non-minimal couplings  to gravity can be determined from
CMB observations and turn out to be highly hierarchical ($\xi_\chi\sim
10^{-3}, \xi_h\sim 10^3-10^5$)
\cite{GarciaBellido:2011de,Bezrukov:2009db}. The scale-invariant
potential $U(\chi,h)$ is given by
\be{jord-pot}
U(\chi,h)=\frac{\lambda}{4}
\left(h^2-\frac{\alpha}{\lambda}\chi^2 \right)^2+\beta \chi^4\,,
\ee
with $\lambda$ the self-coupling of the Higgs field. The parameters
$\alpha$ and $\beta$ must be properly tuned in order to reproduce the
correct hierarchy between the  electroweak, Planck and cosmological
constant scales.  In particular, we must require  $\beta\ll\alpha \ll
1$. The smallness of all the couplings involving the dilaton field
gives rise to an approximate shift symmetry
$\chi\rightarrow\chi+\textrm{const.}$, which, as described in Ref.
\cite{Bezrukov:2010jz}, has important consequences for the analysis
of quantum e\-ffects. For the typical energy scales involved in the
(p)reheating stage we can safely set $\alpha=\beta=0$ in all the
following developments.

We study here the (p)reheating of the universe after Higgs-Dilaton
inflation. As emphasized in Ref. \cite{GarciaBellido:2011de}, particle
production is more easily a\-nalyzed in the so-called Einstein-frame,
where the Higgs and dilaton fields are minimally coupled to gravity.
Performing a conformal redefinition of the metric,  $\tilde
g_{\mu\nu}=\Omega^2g_{\mu\nu}$, with conformal factor 
$\Omega^2=M_P^{-2}(\xi_\chi \chi^2 +\xi_h h^2)$, we obtain
\be{einst-theory}
\frac{\mathscr L}{\sqrt{-\tilde g}}=
\frac{M_P^2}{2}\tilde R-\frac{1}{2}\tilde K(\chi,h)-\tilde U(\chi,h) \ .
\ee
Here $\tilde K(\chi,h)$ is a non-canonical kinetic term in the basis
$(\phi^1, \phi^2)=(\chi,h)$ 
\be{kinetic}
\tilde K(\chi,h) = \frac{\kappa_{ij}}{\Omega^2}
\tilde g^{\mu\nu}\partial_\mu\phi^i\partial_\nu\phi^j\,,
\ee
with
\be{kappa} \kappa_{ij}=\left(\delta_{ij}+
\frac{3}{2}M_P^2\frac{\partial_i\Omega^2\partial_j\Omega^2}
{\Omega^2} \right) \,,
\ee
and $\tilde U(\chi,h)\equiv U(\chi,h)/\Omega^4$ is the Einstein-frame
potential.  In order to diagonalize the kinetic term we can make use
of the conserved Noether's current a\-ssociated to scale invariance. It
can be easily shown, via the homogeneous Friedmann and Klein-Gordon equations in
the slow-roll approximation, that the field combination
$(1+6\xi_\chi)\chi^2+(1+6\xi_h)h^2$ is time-independent in the absence
of any explicit symmetry breaking term \cite{GarciaBellido:2011de}.
This conservation suggests a field redefinition to polar variables in
the $(h,\chi)$ plane
\be{var-rho}
r= \frac{M_P}{2}\log\left[\frac{(1+6\xi_\chi)\chi^2+
(1+6\xi_h)h^2}{M_P^2} \right] \,,
\ee
\be{var-theta}
\tan\theta=\sqrt{\frac{1+6\xi_h}{1+6\xi_\chi}}\frac{h}{\chi} \,.
\ee
In terms of the new coordinates, the kinetic term \eqref{kinetic} becomes dia\-gonal,
although non-canonical,
\ba{polar-kin}
\tilde K &=& \left( \frac{1+6\xi_h}{\xi_h}\right)\frac{1}{\sin^2\theta+\varsigma\cos^2\theta}
(\partial r)^2 \nonumber \\&+&\frac{M_P^2 \ \varsigma}{\xi_\chi}
\frac{\tan^2\theta+\mu}{\cos^2\theta(\tan^2\theta+\varsigma)^2}(\partial \theta)^2 \, ,
\ea
where we have defined
\be{var-defs1}
\mu=\frac{\xi_\chi}{\xi_h} \ \
\ \text{and} \ \ \  \varsigma=\frac{(1+6\xi_h)\xi_\chi}{(1+6\xi_\chi)\xi_h} \ .
\ee 
The dilatonic field $r$ is massless, as corresponds to the Goldstone
boson associated with the spontaneously broken scale symmetry. The
inflationary potential depends only on the angular variable $\theta$
and it is symmetric around $\theta=0$
\be{polar-pot}
\tilde U(\theta)= \frac{\lambda M_P^4}{4\xi_h^2}
\left(\frac{\sin^2\theta}
{\sin^2\theta+\varsigma\cos^2\theta}\right)^2 \ . 
\ee
It can be easily seen that during the (p)reheating stage the values of
the oscillating field $\theta$ are much larger than the $\mu$
parameter for a large number of oscillations,  $\tan^2\theta\gg\mu$.
This allows us  to neglect the $\mu$ term in Eq. \eqref{polar-kin} and
perform an extra field redefinition
\be{var-r-theta}
\rho= \gamma^{-1}r\,,\ \ \ \   
\vert\phi\vert= \phi_0-\frac{M_P}{a}\tanh^{-1}\left[\sqrt{1-\varsigma}\cos\theta\ \right] \,, 
\ee
with
\be{var-defs2}
\gamma= \sqrt{\frac{\xi_\chi}{1+6\xi_\chi}} 
\ \ \ \text{and} \ \ \ a=\sqrt{\frac{\xi_\chi(1-\varsigma)}{\varsigma}} \ .
\ee
The variable $\phi$ is periodic and defined in the compact interval
$\phi \in\left[-\phi_0,\phi_0\right]$, where $\phi_0=
M_P/a\tanh^{-1}\left[\sqrt{1-\varsigma}\ \right] $ corresponds to the
value of the  field $\phi$ at the beginning of inflation. As happened
in Higgs inflation  \cite{Bezrukov:2008ut,GarciaBellido:2008ab}, the
absolute value in Eq.
\eqref{var-r-theta} is required for $\phi$ to maintain the symmetry of the initial $\theta$ field around the minimum of the potential.
In terms of these variables the Lagrangian \eqref{einst-theory} takes a
very simple form
\be{angul-theory}
\frac{\mathscr L}{\sqrt{-\tilde g}}=\frac{M_P^2}{2}\tilde R -
\frac{e^{2b\left(\phi\right)}}{2} (\partial \rho)^2- 
\frac{1}{2}(\partial \phi)^2-\tilde U(\phi)\ ,
\ee
that has been widely studied in the literature \cite{GarciaBellido:1995fz,GarciaBellido:1995qq,DiMarco:2002eb,Lachapelle:2008sy}. In our case, the coefficient in front of the dilaton kinetic term is given by
\be{exp2b}
e^{2b\left(\phi\right)}\equiv\varsigma\cosh^2\left[a \kappa\left(\phi_0-\vert\phi\vert \right)\right]\,,
\ee
with $\kappa$ the inverse of the reduced Planck mass $M_P$.  The inflationary potential \eqref{polar-pot} becomes
\be{defin-pot}
\tilde U(\phi)=\frac{\lambda M_P^4}{4\xi_h^2 (1-\varsigma)^2}
\left(1- \varsigma\cosh^2[a\kappa\left(\phi_0-\vert\phi\vert\right)]\right)^2\,.
\ee
\begin{figure}
\centering
\includegraphics[scale=0.7]{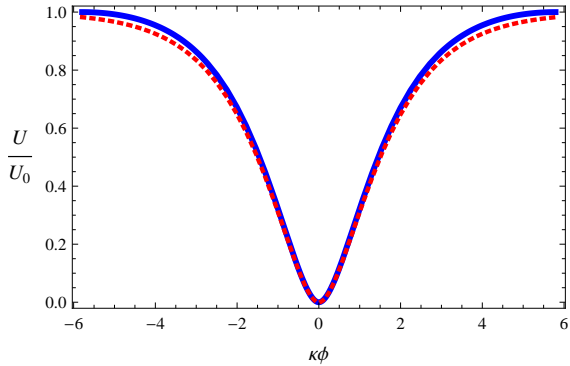}
\caption{Comparison between the Higgs-Dilaton inflatio\-nary potential
(blue continuous line) obtained from Eq.\eqref{defin-pot} and the
corresponding one for the  Higgs inflation mode (red dotted line). In
spite of the slight differences in the upper inflationary region, they
nicely agree in the lower part, where  the (p)reheating stage takes
place. Here $U_0=\lambda M_P^4/(4\xi_h^2)$ and $\kappa=M_P^{-1}$.}
 \label{comparisonHDH}
\end{figure}
%%%%%%%%%%%%%%%%%%%%%%%%%%%%%
\section{SM particle production}\label{sec:gaugesector} 
%%%%%%%%%%%%%%%%%%%%%%%%%%%%%
 As shown in Fig.~\ref{comparisonHDH}, the shape of the Higgs-Dilaton
potential \eqref{defin-pot}  clearly resembles that of the simplest
Higgs inflationary scenario. In spite of the slight differences, both
of them present an exponentially flat region for large field values
and nicely agree for small ones. Indeed, the relation between them
becomes explicit if we approximate Eq. (\ref{defin-pot}) for small
$\phi$. The potential around the minimum behaves, in a good
approximation, as the standard chaotic potential\footnote{It is
interesting to notice at this point that this statement is only valid
for values of the field such that $\tan^2\theta \gg \mu$. For very
small values of the angular variable $\phi$ we recover the standard
$\lambda \phi^4$ Higgs potential. As it happens in Higgs inflation 
\cite{Bezrukov:2008ut,GarciaBellido:2008ab}, this region turns out to
be extremely small, being completely irrelevant for the study of the
(p)reheating stage.}
 \begin{equation}\label{HDpot}
\tilde U(\phi)\simeq \frac{1}{2} M_{HD}^2 \phi^2 + {\cal O}(\phi^3)\,,
\end{equation}
where the higher order corrections can be safely neglected after a few
oscillations.  The curvature of the potential,
$M_{HD}^2=\left(1+6\xi_\chi\right)M^2$, coincides, up to small
corrections, with that of Higgs inflation $M^2=\frac{\lambda
M_P^2}{3\xi_h^2}$. The same applies to the masses of gauge bosons and
fermions.  We obtain  $ (\tilde m_{{\cal A},f}^2)_{HD}=\tilde 
m^2_{{\cal A},f} \left(1+6 \xi_\chi\right)$, with $\tilde m^2_{{\cal
A},f} \simeq \frac{\alpha g^2 M_P}{4\xi_h }|\phi| $ the Einstein-frame
gauge boson and fermion masses in the Higgs inflation model 
\cite{Bezrukov:2008ut,GarciaBellido:2008ab}. Here  $g = g_2,
g_2/\cos\theta_w$ and $\sqrt{2}y_f$, for ${\cal A}=W,Z$ bosons and
fermions $f$, respectively.  We see therefore that,  from the point of
view of (p)reheating, all the relevant physical scales in Higgs and
Higgs-Dilaton inflation coincide, up to small corrections proportional
to the small parameter $\xi_\chi$. This allows us to apply the main
results of Refs. \cite{Bezrukov:2008ut,GarciaBellido:2008ab} to the
Higgs-Dilaton case. Let us summarize here those results. In Higgs
inflation, the SM particles are produced through the so-called
Combined Preheating mechanism 
~\cite{Bezrukov:2008ut,GarciaBellido:2008ab}. Intermediate $W^{\pm}$
and $Z$ bosons are created from the oscillations of the Higgs at the
bottom of the potential~\eqref{HDpot}, whenever there is a violation
of the adiabaticity condition. While there is no restriction on the
number of created gauge bosons, the direct production of SM fermions
by this mechanism is severely res\-tricted by Fermi-Dirac statistics.
The SM fermions appear as se\-condary products of the weak bosons
created in each zero crossing.  Once produced, the gauge bosons
acquire a large effective mass due to the increasing expectation value
of the Higgs field and decay perturbatively into quarks and leptons
within a semioscillation of the Higgs field. This decay rapidly
depletes their occupation numbers and postpones the development of
parametric re\-sonance. During the first oscillations, the fraction of
energy into SM particles is still very small compared with the energy
in the oscillating Higgs field. A large number of oscillations ($
t\sim 300 \,M_{HD}^{-1}$) will be needed in order to transfer a
significant amount of energy into the  SM bosons and fermions. The
decreasing of the Higgs amplitude due to the expansion of the Universe
eventually reduces the decay rate and parametric resonance becomes the
dominant effect.  At this point, the gauge bosons start to build up
their occupation numbers via bosonic stimulation and reheating occurs
within a few osci\-llations. Soon afterwards, the Universe is filled
with the remnant Higgs condensate and a non-thermal distribution of
fermions and bosons, redshifting as radiation and matter respectively.
From there on until thermalization, the evolution of the system is
highly non-linear and non-perturbative, which makes difficult to make
a clear statement about the subsequent evolution without the use of
numerical lattice
simulations~\cite{Bezrukov:2008ut,GarciaBellido:2008ab} .  However, 
thermal equili\-brium is expected to be achieved at a rehea\-ting
temperature $T_0\sim (3-15)\times10^{13}$ GeV, much above the QCD phase
transition scale, $T_\textrm{QCD}\sim 300$ MeV, due to the large SM
couplings  \cite{Bezrukov:2008ut,GarciaBellido:2008ab}. 

%%%%%%%%%%%%%%%%%%%%%%%%%%%%%
\section{Dilaton production}\label{sec:dilatonproc}
%%%%%%%%%%%%%%%%%%%%%%%%%%%%%
In addition to the SM fields, the Higgs-Dilaton inflationary scenario
incorporates an extra degree of freedom, the dilaton field $\rho$. The
constancy of the classical background  component  is of course
gua\-ranteed by the scale invariance current conservation, but this
reasoning does not apply to the corresponding quantum excitations. As
suggested in Ref. \cite{Lachapelle:2008sy}, these modes can be excited
in the preheating stage after inflation\footnote{Any dilaton
production previous  to this stage is completely diluted by the
inflationary expansion.} through the non-canonical kinetic term in the
Einstein-frame Lagrangian \eqref{angul-theory}, which mixes quantum
excitations and background solutions. Although the perturbative
dilaton production through this mixing is expected to be very small,
non-perturbative effects might play an important role
\cite{Lachapelle:2008sy}. In this section we estimate the energy
density residing in the dilaton field at the end of the preheating
stage.  Let us start by considering the linearized\footnote{ The
coupled Higgs-Dilaton equations further simplify since the Higgs
fluctuations are not significantly
amplified~\cite{GarciaBellido:2008ab}, and thus can be treated as
decoupled equations.} equations of motion for dilaton perturbations
$\delta \rho_k$ in Fourier space 
% \cite{GarciaBellido:1995fz,DiMarco:2002eb}
% \begin{equation}\label{perturbations1}
%\delta \ddot\phi_k+3H\delta\dot\phi_k+ \left(\frac{k^2}{a^2}+V_{,\phi\phi}\right)\delta\phi_k=0\,,
 %\end{equation}
 \begin{equation}\label{perturbations}
\delta  \ddot\rho_k+\left(3H+2 \dot b\right)\delta\dot\rho_k+\frac{k^2}{a^2}\delta\rho_k=0\,,
 \end{equation}
where we have ignored metric perturbations and taken into account the
constancy of the background field $\rho$  during the Higgs
oscillations at the end of inflation. The function $b=b(\phi)$  plays
the role of an additional oscillatory damping term for the dilaton
perturbations and depends on the absolute value of the inflaton field
$\phi$, cf. Eq. \eqref{exp2b}. This dependence makes it cumbersome
the direct application of the techniques presented in Ref.
\cite{Lachapelle:2008sy} for the study of particle creation in models
with non-canonical kinetic terms. The field redefinitions used there
would imply delta functions coming from the derivatives of the
absolute value, which substantially complicates the analytic and
numerical treatment of the problem. On the other hand, although the
rephrasing of Eq. \eqref{perturbations} as a Hill's equation is
extremely useful for the understanding of the particle creation
mechanisms, it is not necessary for a precise computation in an
expanding background, as the one needed to estimate $N_\textrm{eff}$.
For this reason, we will adopt an alternative approach, dealing only
with non-singular evolution equations in an expanding Universe. Let us
rewrite Eq. \eqref{perturbations} as
 \begin{equation}\label{perturbations2}
\frac{1}{a^3 e^{2b}}\frac{d}{dt}\left(a^3 e^{2b}\frac{d}{dt}\delta\rho_k\right)+\frac{k^2}{a^2}\delta\rho_k=0\,,
 \end{equation}
which, after a redefinition of time, $d\tau=a^{-3} e^{-2b}dt$, can be
recast in the form of a time-dependent harmonic oscillator
 \be{dilatonconf}
\delta\rho''_k+\omega_k^2(\tau)\delta\rho_k=0\,,
 \ee 
with frequency $\omega_k^2(\tau)=k^2 a^4 e^{4b}$. Here the prime
denotes derivative with respect to the new time $\tau$.  Choosing an
initial vacuum state with zero particle content\footnote{This corresponds to the initial vacuum  conditions $\delta\rho_k(0)=1/\sqrt{2\omega_k}$ and $\delta\rho'_k(0)=-i\omega_k\delta\rho_k$. } , the number of created
dilatons  is given by
\be{nk}
 n_k+\frac{1}{2}=\frac{1}{2\omega_k}\Big(\vert\delta\rho'_k\vert^2+ \omega_k^2 \vert\delta \rho_k\vert^2\Big)\,, 
 \ee
 and its associated energy density,
\be{rhok}
 \hspace{15mm} \rho_\chi=\frac{1}{2\pi^2}\int_0^\infty dk k^2\,\omega_k n_k\,,
 \ee
can be computed numerically by solving Eq. \eqref{dilatonconf},
together with the background evolution equations. The resulting energy
density must be compared with  the energy density in SM particles at
the end of the preheating stage, 
$C\equiv\rho_\chi^0/\rho^0_\textrm{SM}$. This quantity can be easily
related to the effective number of light degrees of freedom
$N_\textrm{eff}$. In order to do that, let us note that, once produced, the
dilaton particles are completely decoupled from the SM particles,
being its energy density only diluted by the expansion of the Universe
$\rho_\chi^0 a_{\textrm{0}}^4=\rho_\chi a_{\textrm{f}}^4$.  Here the
subscripts '0' and 'f' stand for the end of the preheating stage and
the BBN epoch respectively. On the other hand, the total entropy of SM
particles after thermalization remains constant,  $s_{\textrm{0}}
a^3_{\textrm{0}}=s_{\textrm{f}} a^3_{\textrm{f}}$. Taking into account
the relation between the entropy density and the number of
relativistic degrees of freedom $g$ at a given temperature,
$s=\frac{\pi^2}{30}g T^3$, the previous expression can be rewritten as
$g_0 T_0^3 a_0^3= g_f T_f^3 a_f^3$.  By combining this expression with the
evolution equation for the dilaton energy density described above and
dividing the result by the energy density stored in a single neutrino
species, $\rho_{\nu}=\frac{\pi^2}{30}g_{\nu}T_{\textrm{f}}^4$, we get
 \be{Neff}
 \Delta N_\textrm{eff}\equiv\left(\frac{\rho_\chi}{\rho_\nu}\right)_\textrm{f}=\frac{g_0}{g_\nu}\left(\frac{g_f}{g_0}\right)^{4/3}C\simeq 2.85\, C\,.
 \ee
In the last equality we have made use of the number of SM degrees of
freedom at the end of inflation ($g_{\textrm{0}}=106.75$) and at BBN
($g_{\textrm{f}}=10.75$). Therefore, we see that, in order to have a
contribution to the effective number of relativistic degrees of
freedom within the reach of the Planck satellite, roughly a $10 \%$ of
the energy density at the end of inflation must be converted into
dilatons. Nevertheless, the transferred fraction turns out to be
signi\-ficantly smaller. Evaluating the numerical solution of
Eq.~\eqref{dilatonconf} at the time at which the energy density in SM
particles roughly equals the initial energy density of the inflaton
field, $\tau(t_0 \sim 300 \,M_{HD}^{-1})$ 
\cite{GarciaBellido:2008ab}, we get $C\sim 10^{-7}$.  The precise
value of the $C$ parameter weakly depends on the ratio
$\xi_h/\sqrt{\lambda}$, which determines the total energy density
available at the end of inflation \cite{GarciaBellido:2011de}, and it is 
quite insensitive to the particular value of the small non-minimal coupling
 $\xi_\chi$. 

Although the non-perturbative creation of dilatons due
to the background field $\phi$ turns out to be extremely small, one should consider
 the possi\-bility of producing them as secondary products of the Higgs particles
  created at the preheating and thermalization stages. The Lagrangian
  \eqref{angul-theory} leads  to a number of perturbative processes, such as the decay of Higgs particles 
into dilatons ($\phi\to\rho\rho$) or Higgs-Higgs scatterings ($\phi\phi\to\rho\rho$). The
corresponding decay rate and cross-section are given respectively by
\be{gammasigma}
\Gamma \simeq \frac{M_{HD}^3}{192\pi M_P^2} \ \ \ \ \text{and} \ \ \ \ 
\sigma \simeq \frac{E^2}{576\pi M_P^4}~,
\ee
where $M_\textrm{HD}^2$ is the effective Higgs mass in the region  $M_P/\xi_h < \phi \ll M_P$ (cf. Eq.~\eqref{HDpot}) and $E$ is the Higgs energy in the center-of-mass frame. Assuming this energy to be of the order of the temperature of the thermalized SM plasma, $T_0$, it can be easily seen that the contribution of these processes to the $C$ parameter is of order
\be{cpert}
C \propto \left(\frac{T_0}{M_P}\right)^3\,,
\ee
and therefore much smaller than the non-perturbative contribution found above. 

%%%%%%%%%%%%%%%%%%%%%%%%%%%%%
\section{Conclusions}\label{sec:conclusions}
%%%%%%%%%%%%%%%%%%%%%%%%%%%%%
We have considered particle production in a scale-invariant extension
of Higgs inflation known as Higgs-Dilaton inflation. This model
predicts the existence of an extra massless particle -  the dilaton -
which might contribute to the effective number of light degrees of
freedom. After recasting the problem in the appropriate set of
varia\-bles, all the particle masses and energy scales in the model
turn out to coincide, up to small corrections, with those of Higgs
inflation. Gauge bosons and fermions are therefore produced through
the so-called Combined Preheating mechanism.  On the other hand, the
production of dilatons could take place only through the non-canonical
kinetic term for the dilaton field, fed by the absolute value of the
Higgs field. We have shown that the number of non-perturbatively
created particles can be easily eva\-luated thanks to a particular time
redefinition. The dilaton energy density computed this way turns out
to be extremely small, which is translated into an effective number of
relativistic degrees of freedom  very close to the SM one. This result
is not modified by any of the subsequent perturbative processes
involving the Higgs particles produced at the bottom of the potential.
We thus conclude that, in spite of the potential value of BBN and CMB
for testing the existence of the elusive dilaton particle, its
subdominant production at the (p)reheating stage after inflation
makes  it completely undetectable by any particle physics experiment
or cosmological observation. The only remnant of the dilaton field is
a dynamical Dark Energy stage with an equation of state close, but slightly
different, to that of a cosmological constant, which leads to a
power-like expansion of the Universe in the far future
\cite{Shaposhnikov:2008xb}.

%%%%%%%%%%%%%%%%%%%%%%%%%%%%%
\section*{Acknowledgements}
%%%%%%%%%%%%%%%%%%%%%%%%%%%%%
We thank Daniel Zenh\"ausern for collaboration in the early stages of this project and Julien Lesgourgues for useful discussions. 
This work was partially supported by the Swiss National Science
Foundation. We also acknowledge  financial support from the Madrid
Regional Government (CAM) under the program  HEPHACOS
S2009/ESP-1473-02, from MICINN under grant  AYA2009-13936-C06-06  and
Consolider-Ingenio 2010 PAU (CSD2007-00060), as well as from the
European  Union Marie Curie Initial Training Network "UNILHC"
PITN-GA-2009-237920.


\begin{thebibliography}{99}

%\cite{Steigman:1977kc}
\bibitem{Steigman:1977kc}
  G.~Steigman, D.~N.~Schramm and J.~E.~Gunn,
  %``Cosmological Limits to the Number of Massive Leptons,''
  Phys.\ Lett.\ B {\bf 66} (1977) 202.
  %%CITATION = PHLTA,B66,202;%%
  
%\cite{Beringer:1900zz}
\bibitem{Beringer:1900zz}
  J.~Beringer {\it et al.}  [Particle Data Group Collaboration],
  %``Review of Particle Physics (RPP),''
  Phys.\ Rev.\ D {\bf 86} (2012) 010001.
  %%CITATION = PHRVA,D86,010001;%%


%\cite{Lesgourgues:2006nd}
\bibitem{Lesgourgues:2006nd}
  J.~Lesgourgues and S.~Pastor,
  %``Massive neutrinos and cosmology,''
  Phys.\ Rept.\  {\bf 429} (2006) 307
  [astro-ph/0603494].
  %%CITATION = ASTRO-PH/0603494;%%
  
  %\cite{Steigman:2012ve}
\bibitem{Steigman:2012ve}
  G.~Steigman,
  %``Neutrinos And Big Bang Nucleosynthesis,''
  arXiv:1208.0032 [hep-ph].
  %%CITATION = ARXIV:1208.0032;%%

%\cite{Komatsu:2010fb}
 \bibitem{Komatsu:2010fb}
  E.~Komatsu {\it et al.}  [WMAP Collaboration],
  %``Seven-Year Wilkinson Microwave Anisotropy Probe (WMAP) Observations: Cosmological Interpretation,''
  Astrophys.\ J.\ Suppl.\  {\bf 192} (2011) 18
  [arXiv:1001.4538 [astro-ph.CO]].
  %%CITATION = ARXIV:1001.4538;%% 
  
%\cite{Dunkley:2010ge}
\bibitem{Dunkley:2010ge}
  J.~Dunkley, R.~Hlozek, J.~Sievers, V.~Acquaviva, P.~A.~R.~Ade, P.~Aguirre, M.~Amiri and J.~W.~Appel {\it et al.},
  %``The Atacama Cosmology Telescope: Cosmological Parameters from the 2008 Power Spectra,''
  Astrophys.\ J.\  {\bf 739} (2011) 52
  [arXiv:1009.0866 [astro-ph.CO]].
  %%CITATION = ARXIV:1009.0866;%%

%\cite{Keisler:2011aw}
\bibitem{Keisler:2011aw}
  R.~Keisler, C.~L.~Reichardt, K.~A.~Aird, B.~A.~Benson, L.~E.~Bleem, J.~E.~Carlstrom, C.~L.~Chang and H.~M.~Cho {\it et al.},
  %``A Measurement of the Damping Tail of the Cosmic Microwave Background Power Spectrum with the South Pole Telescope,''
  Astrophys.\ J.\  {\bf 743} (2011) 28
  [arXiv:1105.3182 [astro-ph.CO]].
  %%CITATION = ARXIV:1105.3182;%%

%\cite{GonzalezGarcia:2010un}
\bibitem{GonzalezGarcia:2010un}
  M.~C.~Gonzalez-Garcia, M.~Maltoni and J.~Salvado,
  %``Robust Cosmological Bounds on Neutrinos and their Combination with Oscillation Results,''
  JHEP {\bf 1008} (2010) 117
  [arXiv:1006.3795 [hep-ph]].
  %%CITATION = ARXIV:1006.3795;%%

%\cite{Hou:2011ec}
\bibitem{Hou:2011ec}
  Z.~Hou, R.~Keisler, L.~Knox, M.~Millea and C.~Reichardt,
  %``How Massless Neutrinos Affect the Cosmic Microwave Background Damping Tail,''
  arXiv:1104.2333 [astro-ph.CO].
  %%CITATION = ARXIV:1104.2333;%%

%\cite{Reid:2009nq}
\bibitem{Reid:2009nq}
  B.~A.~Reid, L.~Verde, R.~Jimenez and O.~Mena,
  %``Robust Neutrino Constraints by Combining Low Redshift Observations with the CMB,''
  JCAP {\bf 1001} (2010) 003
  [arXiv:0910.0008 [astro-ph.CO]].
  %%CITATION = ARXIV:0910.0008;%%

%\cite{GonzalezMorales:2011ty}
%\bibitem{GonzalezMorales:2011ty}
 % A.~X.~Gonzalez-Morales, R.~Poltis, B.~D.~Sherwin and L.~Verde,
  %``Are priors responsible for cosmology favoring additional neutrino species?,''
 % arXiv:1106.5052 [astro-ph.CO].
  %%CITATION = ARXIV:1106.5052;%%

%\cite{Pettini:2012ph}
\bibitem{Pettini:2012ph}
  M.~Pettini and R.~Cooke,
  %``A new, precise measurement of the primordial abundance of Deuterium,''
  arXiv:1205.3785 [astro-ph.CO].
  %%CITATION = ARXIV:1205.3785;%%

%\cite{Ichikawa:2008pz}
\bibitem{Ichikawa:2008pz}
  K.~Ichikawa, T.~Sekiguchi and T.~Takahashi,
  %``Probing the Effective Number of Neutrino Species with Cosmic Microwave Background,''
  Phys.\ Rev.\ D {\bf 78} (2008) 083526
  [arXiv:0803.0889 [astro-ph]].
  %%CITATION = ARXIV:0803.0889;%%

%\cite{Hamann:2010pw}
\bibitem{Hamann:2010pw}
  J.~Hamann, S.~Hannestad, J.~Lesgourgues, C.~Rampf and Y.~Y.~Y.~Wong,
  %``Cosmological parameters from large scale structure - geometric versus shape information,''
  JCAP {\bf 1007} (2010) 022
  [arXiv:1003.3999 [astro-ph.CO]].
  %%CITATION = ARXIV:1003.3999;%%

%\cite{Hamann:2010bk}
\bibitem{Hamann:2010bk}
  J.~Hamann, S.~Hannestad, G.~G.~Raffelt, I.~Tamborra and Y.~Y.~Y.~Wong,
  %``Cosmology seeking friendship with sterile neutrinos,''
  Phys.\ Rev.\ Lett.\  {\bf 105} (2010) 181301
  [arXiv:1006.5276 [hep-ph]].
  %%CITATION = ARXIV:1006.5276;%%
  
%\cite{Hannestad:2012ky}
\bibitem{Hannestad:2012ky}
  S.~Hannestad, I.~Tamborra and T.~Tram,
  %``Thermalisation of light sterile neutrinos in the early universe,''
  JCAP {\bf 1207} (2012) 025
  [arXiv:1204.5861 [astro-ph.CO]].
  %%CITATION = ARXIV:1204.5861;%%

%\cite{Giovannini:2002qw}
\bibitem{Giovannini:2002qw}
  M.~Giovannini, H.~Kurki-Suonio and E.~Sihvola,
  %``Big bang nucleosynthesis, matter antimatter regions, extra relativistic species, and relic gravitational waves,''
  Phys.\ Rev.\ D {\bf 66} (2002) 043504
  [astro-ph/0203430].
  %%CITATION = ASTRO-PH/0203430;%%

%\cite{Ichikawa:2007jv}
\bibitem{Ichikawa:2007jv}
  K.~Ichikawa, M.~Kawasaki, K.~Nakayama, M.~Senami and F.~Takahashi,
  %``Increasing effective number of neutrinos by decaying particles,''
  JCAP {\bf 0705} (2007) 008
  [hep-ph/0703034 [HEP-PH]].
  %%CITATION = HEP-PH/0703034;%%


%\cite{Zhang:2007zzh}
\bibitem{Zhang:2007zzh}
  L.~Zhang, X.~Chen, M.~Kamionkowski, Z.~-g.~Si and Z.~Zheng,
  %``Constraints on radiative dark-matter decay from the cosmic microwave background,''
  Phys.\ Rev.\ D {\bf 76} (2007) 061301
  [arXiv:0704.2444 [astro-ph]].
  %%CITATION = ARXIV:0704.2444;%%

%\cite{Fischler:2010xz}
\bibitem{Fischler:2010xz}
  W.~Fischler and J.~Meyers,
  %``Dark Radiation Emerging After Big Bang Nucleosynthesis?,''
  Phys.\ Rev.\ D {\bf 83} (2011) 063520
  [arXiv:1011.3501 [astro-ph.CO]].
  %%CITATION = ARXIV:1011.3501;%%

%\cite{Bjaelde:2012wi}
\bibitem{Bjaelde:2012wi}
  O.~E.~Bjaelde, S.~Das and A.~Moss,
  %``Origin of Delta N_eff as a Result of an Interaction between Dark Radiation and Dark Matter,''
  arXiv:1205.0553 [astro-ph.CO].
  %%CITATION = ARXIV:1205.0553;%%

%\cite{Boehm:2012gr}
\bibitem{Boehm:2012gr}
  C.~Boehm, M.~J.~Dolan and C.~McCabe,
  %``Increasing Neff with particles in thermal equilibrium with neutrinos,''
  arXiv:1207.0497 [astro-ph.CO].
  %%CITATION = ARXIV:1207.0497;%%

%\cite{Shaposhnikov:2008xb}
 \bibitem{Shaposhnikov:2008xb}
  M.~Shaposhnikov and D.~Zenhausern,
  %``Scale invariance, unimodular gravity and dark energy,''
  Phys.\ Lett.\ B {\bf 671} (2009) 187
  [arXiv:0809.3395 [hep-th]].
  %%CITATION = ARXIV:0809.3395;%%

%\cite{GarciaBellido:2011de}
 \bibitem{GarciaBellido:2011de}
  J.~Garcia-Bellido, J.~Rubio, M.~Shaposhnikov and D.~Zenhausern,
  %``Higgs-Dilaton Cosmology: From the Early to the Late Universe,''
  Phys.\ Rev.\ D {\bf 84} (2011) 123504
  [arXiv:1107.2163 [hep-ph]].
  %%CITATION = ARXIV:1107.2163;%%
  
  %\cite{Shaposhnikov:2008xi}
\bibitem{Shaposhnikov:2008xi}
  M.~Shaposhnikov and D.~Zenhausern,
  %``Quantum scale invariance, cosmological constant and hierarchy problem,''
  Phys.\ Lett.\ B {\bf 671} (2009) 162
  [arXiv:0809.3406 [hep-th]].
  %%CITATION = ARXIV:0809.3406;%%
  
  %\cite{Bezrukov:2007ep}
 \bibitem{Bezrukov:2007ep}
  F.~L.~Bezrukov and M.~Shaposhnikov,
  %``The Standard Model Higgs boson as the inflaton,''
  Phys.\ Lett.\ B {\bf 659} (2008) 703
  [arXiv:0710.3755 [hep-th]].
  %%CITATION = ARXIV:0710.3755;%%

%\cite{Bezrukov:2008ut}
 \bibitem{Bezrukov:2008ut}
  F.~Bezrukov, D.~Gorbunov and M.~Shaposhnikov,
  %``On initial conditions for the Hot Big Bang,''
  JCAP {\bf 0906} (2009) 029
  [arXiv:0812.3622 [hep-ph]].
  %%CITATION = ARXIV:0812.3622;%%
  
%\cite{GarciaBellido:2008ab}
 \bibitem{GarciaBellido:2008ab}
  J.~Garcia-Bellido, D.~G.~Figueroa and J.~Rubio,
  %``Preheating in the Standard Model with the Higgs-Inflaton coupled to gravity,''
  Phys.\ Rev.\ D {\bf 79} (2009) 063531
  [arXiv:0812.4624 [hep-ph]].
  %%CITATION = ARXIV:0812.4624;%%

%\cite{Bezrukov:2011sz}
\bibitem{Bezrukov:2011sz}
  F.~Bezrukov, D.~Gorbunov and M.~Shaposhnikov,
  %``Late and early time phenomenology of Higgs-dependent cutoff,''
  JCAP {\bf 1110} (2011) 001
  [arXiv:1106.5019 [hep-ph]].
  %%CITATION = ARXIV:1106.5019;%%
  
  %\cite{Bezrukov:2009db}
\bibitem{Bezrukov:2009db}
  F.~Bezrukov and M.~Shaposhnikov,
  %``Standard Model Higgs boson mass from inflation: Two loop analysis,''
  JHEP {\bf 0907} (2009) 089
  [arXiv:0904.1537 [hep-ph]].
  %%CITATION = ARXIV:0904.1537;%%
  
  %\cite{Bezrukov:2010jz}
\bibitem{Bezrukov:2010jz}
  F.~Bezrukov, A.~Magnin, M.~Shaposhnikov and S.~Sibiryakov,
  %``Higgs inflation: consistency and generalisations,''
  JHEP {\bf 1101} (2011) 016
  [arXiv:1008.5157 [hep-ph]].
  %%CITATION = ARXIV:1008.5157;%%  

%\cite{GarciaBellido:1995fz}
\bibitem{GarciaBellido:1995fz}
  J.~Garcia-Bellido and D.~Wands,
  %``Constraints from inflation on scalar - tensor gravity theories,''
  Phys.\ Rev.\ D {\bf 52} (1995) 6739
  [gr-qc/9506050].
  %%CITATION = GR-QC/9506050;%%
  
  
  %\cite{GarciaBellido:1995qq}
\bibitem{GarciaBellido:1995qq}
  J.~Garcia-Bellido and D.~Wands,
  %``Metric perturbations in two field inflation,''
  Phys.\ Rev.\ D {\bf 53} (1996) 5437
  [astro-ph/9511029].
  %%CITATION = ASTRO-PH/9511029;%%

%\cite{DiMarco:2002eb}
\bibitem{DiMarco:2002eb}
  F.~Di Marco, F.~Finelli and R.~Brandenberger,
  %``Adiabatic and isocurvature perturbations for multifield generalized Einstein models,''
  Phys.\ Rev.\ D {\bf 67} (2003) 063512
  [astro-ph/0211276].
  %%CITATION = ASTRO-PH/0211276;%%

%\cite{Lachapelle:2008sy}
\bibitem{Lachapelle:2008sy}
  J.~Lachapelle and R.~H.~Brandenberger,
  %``Preheating with Non-Standard Kinetic Term,''
  JCAP {\bf 0904} (2009) 020
  [arXiv:0808.0936 [hep-th]].
  %%CITATION = ARXIV:0808.0936;%%







%%%%%%%%%%%%%%%%%%%%%%%%%%%%%%%%%%%%%%%%



\end{thebibliography}
\end{document}